\documentclass[aps,pra,showpacs,twocolumn,superscriptaddress]{revtex4-1}
\usepackage{graphicx}
\usepackage[usenames]{color}
\usepackage{amssymb,amsmath}
\usepackage{siunitx}%To use degree symbol
\usepackage{xcolor}
\usepackage{setspace} 
\usepackage{float} 
\usepackage{tabularx}

\newcommand{\fig}[1]{Fig.~\ref{#1}}
\newcommand{\be}[1]{\begin{equation}\label{#1}}
\newcommand{\ee}{\end{equation}}

\usepackage[LGRgreek]{mathastext}
\usepackage{lipsum}

\begin{document}

\title{Slingshot non-sequential double ionization as a gate to anti-correlated two electron escape}

\author{G. P. Katsoulis}
\affiliation{Department of Physics and Astronomy, University College London, Gower Street, London WC1E 6BT, United Kingdom}
\author{A. Hadjipittas}
\affiliation{Department of Physics and Astronomy, University College London, Gower Street, London WC1E 6BT, United Kingdom}
\author{B. Bergues}
\affiliation{Department of Physics, Ludwig-Maximilians-Universit\"{a}t Munich, Am Coulombwall 1, D-85748 Garching, Germany}
\affiliation{Max Planck Institute of Quantum Optics, Hans-Kopfermann-Str. 1, D-85748 Garching, Germany}
\author{M. F. Kling}
\affiliation{Department of Physics, Ludwig-Maximilians-Universit\"{a}t Munich, Am Coulombwall 1, D-85748 Garching, Germany}
\affiliation{Max Planck Institute of Quantum Optics, Hans-Kopfermann-Str. 1, D-85748 Garching, Germany}
\author{A. Emmanouilidou}
\affiliation{Department of Physics and Astronomy, University College London, Gower Street, London WC1E 6BT, United Kingdom}

%\author{A. Chen}
%\affiliation{Department of Physics and Astronomy, University College London, Gower Street, London WC1E 6BT, United Kingdom}
%\author{C. Lazarou}
%\affiliation{Department of Physics and Astronomy, University College London, Gower Street, London WC1E 6BT, United Kingdom}
%\author{H. Price}
%\affiliation{Department of Physics and Astronomy, University College London, Gower Street, London WC1E 6BT, United Kingdom}
%\author{A. Staudte}
%\affiliation{Joint Laboratory for Attosecond Science, University of Ottawa and National Research Council,
%100 Sussex Drive, Ottawa, Ontario, Canada K1A 0R6}
%\author{I. Ben-Itzhak}
%\affiliation{Kansas State University, Manhattan}

%\author{A. Chen}
%\affiliation{Department of Physics and Astronomy, University College London, Gower Street, London WC1E 6BT, United Kingdom}
%\author{M. F. Kling}
%\affiliation{Department of Physics, Ludwig-Maximilians-Universit\"{a}t Munich, Am Coulombwall 1, D-85748 Garching, Germany}
%\affiliation{Max Planck Institute of Quantum Optics, Hans-Kopfermann-Str. 1, D-85748 Garching, Germany}
%\author{A. Emmanouilidou}
%\affiliation{Department of Physics and Astronomy, University College London, Gower Street, London WC1E 6BT, United Kingdom}

\begin{abstract}
At intensities below-the-recollision threshold,  we show that re-collision-induced  excitation with one electron escaping fast after re-collision and the other electron escaping with a time delay via a Coulomb slingshot motion is one of the most important mechanisms of non-sequential double ionization, for strongly-driven He at 400 nm.  %In this slingshot-NSDI mechanism the electron that is still bound after re-collision  is assisted by the nucleus and the laser field to ionize mostly around the second  extremum of the laser field after re-collision. 
Slingshot-NSDI is a general mechanism present for a wide range of  low intensities  and pulse durations. Anti-correlated two-electron escape   is its striking hallmark. This mechanism offers an alternative explanation 
of  anti-correlated two-electron escape obtained in previous studies.

 \end{abstract}
\date{\today}

\maketitle

Non-sequential double ionization (NSDI) in strong infrared  laser fields is a fundamental process  \cite{Corkum,PhysRevA.27.2503,PhysRevLett.55.2141,PhysRevLett.73.1227,0953-4075-31-6-008,PhysRevLett.84.447,RESI1,WeberNature,RESI2,DORNER20021,kinematic1,PhysRevLett.92.173001,PhysRevLett.93.253001,PhysRevLett.94.093002,NSDI2,Taylor1,vshape1,vshape2,NSDI1,PfeifferNature,Dorner2013,Chen2016,Dorner2018} accounted for
by the  three-step model \cite{Corkum}. First, one electron  tunnel-ionizes in the field-lowered Coulomb-barrier and then accelerates in the  laser field. This electron can return back to the core to re-collide and transfer energy to  a bound electron through    different pathways. In the direct one, the energy  transferred suffices for both electrons to   ionize shortly after re-collision. In the delayed pathway the energy  transferred  ionizes one of the two electrons shortly after re-collision. The other electron transitions to an excited state and ionizes later. 
 It is generally accepted that re-collision-induced excitation  with subsequent field ionization (RESI)  \cite{RESI1,RESI2} prevails the delayed pathway. In RESI, the  excited electron ionizes at the field extrema,  after re-collision, assisted by the laser field.
  In the double delayed pathway,  both electrons ionize later following the energy  transferred during re-collision.

At intensities below the recollision threshold, in NSDI two electrons escaping opposite to each other along the  laser-field direction---anti-correlated escape---has been studied intensely by experiment and theory alike. This pattern  was found to prevail, but not substantially,   over correlated two-electron escape. It  was observed in NSDI  of several  atoms driven by intense (strongly-driven) long duration pulses \cite{Anticor1,Anticor2,Anticor4,Anticor8,Anticor7,Anticor3,Anticor5,Anticor6}. Multiple re-collisions, in the context of  RESI,  were put forth to explain anti-correlated two-electron escape \cite{Anticor1, Anticor2, Anticor4,Anticor8, Anticor7}. Electron-electron repulsion was also suggested as a possible explanation \cite{Anticor5, Anticor6}.    

Here, we show that RESI does not necessarily  prevail the delayed pathway, for strongly-driven He at  400 nm.  We find a competing mechanism in the delayed pathway where the electron that ionizes later undergoes a slingshot motion due to the Coulomb interaction with the nucleus and the field. This Coulomb slingshot motion is similar to the  well known gravitational slingshot that alters the motion of a spacecraft around a planet. Moreover, we find that the electron undergoing a slingshot motion  ionizes around the second extremum  of the laser field after re-collision. We label this mechanism slingshot-NSDI. The  nucleus has a small effect  on RESI. In contrast, in slingshot-NSDI the nucleus plays a decisive role with anti-correlated two-electron escape being its striking hallmark.  Slingshot-NSDI   is an  alternative  to the multiple  re-collisions mechanism for explaining anti-correlated   two-electron escape.

%This key  signature  of RESNI is    particularly pronounced   for laser pulses around  5$\times$10$^{14}$ W/cm$^2$ and of near-single-cycle duration.  In the last few years kinematically complete experiments have been accessing NSDI using carrier-envelope-phase (CEP)-controlled near-single-cycle pulses  \cite{Camus,Kling1,PhysRevA.83.013412}.  For such pulses, 
 %RESNI prevails  the delayed pathway as well as NSDI. Anti-correlated two electron escape overshadows the correlated one. 
  %Moreover,  and that experiments should  find evidence of RESNI in the near future. 
  
  %For near-single-cycle pulses, with increasing intensity the contributions to NSDI of the direct and delayed pathway become comparable. However, 
 %the contribution to the delayed pathway of RESNI and of anti-correlated RESNI events   reduces slowly. For a fixed intensity, with increasing laser-pulse duration  the contributions to NSDI  of the double delayed and delayed pathway become comparable.
%However, the contribution  to the delayed pathway of RESNI  increases, since it can occur  shortly past more even extrema  after re-collision besides the first one.  The contribution  to the delayed pathway of anti-correlated RESNI events    decreases. The larger values of the  laser field  at the second extremum after re-collision reduces the effect of the nucleus  compared to short pulses. 

 We demonstrate slingshot-NSDI in He driven by  a near-single-cycle laser pulse at 5$\times$10$^{14}$ W/cm$^2$ and 400 nm. Further below, we discuss slingshot-NSDI for other laser  intensities and pulse durations  as well.
 Kinematically complete experiments  that employ carrier envelope phase (CEP)-controlled near-single-cycle pulses   have been carried out for NSDI over a wide range of intensities \cite{Anticor7,PhysRevA.83.013412,Camus,Kling1}. 
    The intensity of 5$\times$10$^{14}$ W/cm$^2$ is  below the recollision threshold. This corresponds to  the maximum energy of the  electron returning to the core, 3.17 $\mathcal{E}_{0}^2/(4\omega^2)$ \cite{Corkum}, which is equal to 23.7 eV at 5$\times$10$^{14}$ W/cm$^2$, being equal to the energy   needed to transition  to the first excited state of the ion;   $\mathcal{E}_{0}$ and $\omega$  are the strength and  frequency of the field. 
 
 We employ a three-dimensional (3D) semi-classical model \cite{Agapi_paper_2008,Agapi2,Agapi10}.   One electron (re-colliding) tunnel-ionizes  through the field-lowered Coulomb-barrier.  We use the quantum mechanical Ammosov-Delone-Krainov (ADK) formula to compute the tunnel-ionization rate \cite{A1,A2}.   The exit point of the re-colliding electron  is along the laser-field direction  and is computed using parabolic coordinates \cite{parabolic1}.  The electron momentum   is taken to be equal to zero  along the laser field while  the transverse one  is given by a Gaussian distribution  \cite{A1,A2}.  The initially bound electron  is  described by a microcanonical distribution \cite{Abrimes}.  

   %Also, for intensities above-the-recollision threshold, we showed that electron backscattering  from the nucleus during re-collision gives rise to the the fingerlike structure in the correlated momenta of the two escaping electrons    \cite{Agapi1,vshape3}. This structure was predicted theoretically \cite{Taylor1}, and observed experimentally  \cite{vshape1,vshape2}.

We use a laser field of the form 
\begin{equation}
  \vec{\mathcal{E}}(t) = \mathcal{E}_0 \exp\left(-2\ln 2 \left(\frac{t}{\tau}\right)^2 \right)\cos\left( \omega t + \phi \right) \hat{z}, 
\label{eq1}
 \end{equation}
 where  $\phi$ is the  CEP, and $\tau=2$ fs is the full-width-half-maximum of the pulse duration in intensity. We employ atomic units, unless otherwise stated. The tunnel-ionization time, t$_{0}$,  is selected randomly  in the time interval  [-2$\tau$,2$\tau$].
 Once the initial conditions are specified at time $\mathrm{t_{0}}$, the position and momentum of each electron are propagated classically in time. We do so using  the three-body Hamiltonian of the two electrons with the nucleus kept fixed. All Coulomb forces and the interaction of each electron with the  laser field are fully accounted for with no approximation. We also account for    the Coulomb singularity by using  regularized coordinates \cite{KS}. 
   Our results  for NSDI  are obtained by taking into account  CEPs that range from  $\phi=0^{\circ}$ to $\phi=330^{\circ}$ in steps of 30$^{\circ}$ and by averaging over these CEPs. 
   
   We identify
the main pathways of energy transfer in each double ionization (DI) event.  
To do so, we compute  the  time difference between the re-collision time $\mathrm{t_{rec}}$ and the ionization time of each electron. We compare it with the time interval t$_{diff}$ where the electron pair potential energy undergoes a sharp change due to re-collision. We find t$_{diff}$ to be  roughly equal to 1/8 laser cycle (T).  We refer to  the electron that, after re-collision,  ionizes first as electron 1, and the one ionizing last as  electron 2.
For each classical trajectory, the re-collision time is defined as the time of minimum approach of the two electrons. It is identified by the maximum in the electron pair potential energy. The ionization time for each electron,  $\mathrm{t_{i}}$, is defined as the time when the compensated energy $(p_{x,i}^{2}+p_{y,i}^2+(p_{z,i}-\mathcal{A}(t))^2)/2-Z/r_{i}$ becomes positive and remains positive thereafter \cite{Leopold}, with  $\mathrm{i=1,2}$ and $\mathrm{{\bf{p}}_{i}=p_{x,i}\hat{x}+p_{y,i}\hat{y}+p_{z,i}\hat{z}}$; $\mathcal{A}(\mathrm{t})$ is the vector potential and $Z=2$. %Thus, the ionization time of  electron 1 is not necessarily the time $\mathrm{t_{0}}$ this electron tunnel-ionizes.  
We list the conditions for   direct, delayed or double delayed DI event in Table I. In the delayed pathway,   the probability for electron 2 to be the re-colliding electron  increases with  
decreasing intensity.

 \begin{table}[h]
  $\Delta \mathrm{t_1}$=$\mathrm{t_{1}}$-$\mathrm{t_{rec}}$ \& $\Delta \mathrm{t_2}$=$\mathrm{t_{2}}$-$\mathrm{ t_{rec}}$ \\
  \vspace{0.2cm}
\begin{tabularx}{\columnwidth}{|X|X|X|}
 %\hline
 %\multicolumn{3}{|c|}{$\Delta \mathrm{t_1}$=$\mathrm{t_{1}}$-$\mathrm{t_{rec}}$ \& $\Delta \mathrm{t_2}$=$\mathrm{t_{2}}$-$\mathrm{ t_{rec}}$ } \\
  %\hline
 \hline
 \centering direct & \centering delayed & \centering double delayed \tabularnewline

 \hline
 \centering
$\begin{aligned}
\hspace{0.1cm} \Delta \mathrm{t_2} &< \mathrm{t_{diff}} \\
\hspace{0.1cm}\mathrm{t_1} &<\mathrm{t_2}
\end{aligned}
$

& 
\centering
$\begin{aligned}
\hspace{0.1cm}\Delta \mathrm{t_{1}} &< \mathrm{t_{diff}} \\ 
\hspace{0.1cm}\Delta \mathrm{t_{2}} &>\mathrm{t_{diff}}
\end{aligned}
 $
& 
\centering
$\begin{aligned}
\hspace{0.2cm}\Delta \mathrm{t_{1}} &> \mathrm{t_{diff}}\\ 
\hspace{0.2cm}\Delta \mathrm{t_{2}} &> \mathrm{t_{diff}}
\end{aligned}$ \tabularnewline

\hline

\end{tabularx}
 \caption{Conditions for energy transfer DI pathways. }

\label{Table2}

\end{table}

  At intensities  above-the-recollision threshold the direct pathway dominates.  The two electrons escape mostly in the same direction along the laser field, with  the final momentum of each electron being roughly equal to $-\mathcal{A}(\mathrm{t_{rec}})$. 
  
\begin{figure}[h]
\includegraphics[scale=0.48]{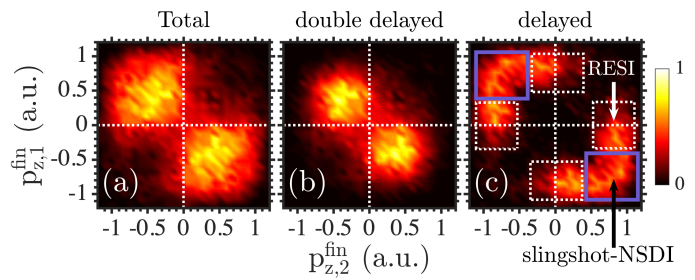}
\centering
\caption{Symmetrized correlated electron momenta  normalized to peak value  for (a) all DI  (b)  double delayed and (c) delayed  events for He driven by a near-single-cycle pulse at 5$\times$10$^{14}$ W/cm$^2$ and 400 nm.  }
\label{fig:cor}
\end{figure}

  At intensities well below-the-recollision threshold the delayed and double delayed pathways prevail. This is the case for He driven by a near-single-cycle  laser pulse at 5$\times$10$^{14}$ W/cm$^2$ and 400 nm. We find that 
   the two electrons  escape overwhelmingly in opposite directions, see \fig{fig:cor}.  In the double delayed  pathway the two electrons escape with small momenta, see \fig{fig:cor}(b), whereas in the delayed pathway the momenta are larger  (\fig{fig:cor}(c)).  To understand the anti-correlation pattern, we focus on the delayed pathway and its contributing mechanisms. 
   
   In the delayed pathway, electron 1 escapes with momentum roughly equal to $-\mathcal{A}(\mathrm{t_{1}\approx t_{rec}})$ which is mostly large,  since  re-collision occurs  around a zero of the  field.
  In RESI,  
  electron 2 ionizes  at a time t$_{2}$ around an extremum of the  field. It does so  primarily due to the laser field, and, thus, has a small final momentum, $-\mathcal{A}(\mathrm{t_{2}})$, along the $\hat{z}$-axis, see  \fig{fig:cor}(c). In the new mechanism, which we refer to as slingshot-NSDI,  the final momentum of electron 2 is determined both by the Coulomb field and the laser field and it has  mostly large magnitude, see \fig{fig:cor}(c).
   Moreover, unlike RESI, in slingshot-NSDI 
   the two electrons escape only in opposite directions.

  \begin{figure*}[ht]
\includegraphics[width=\linewidth]{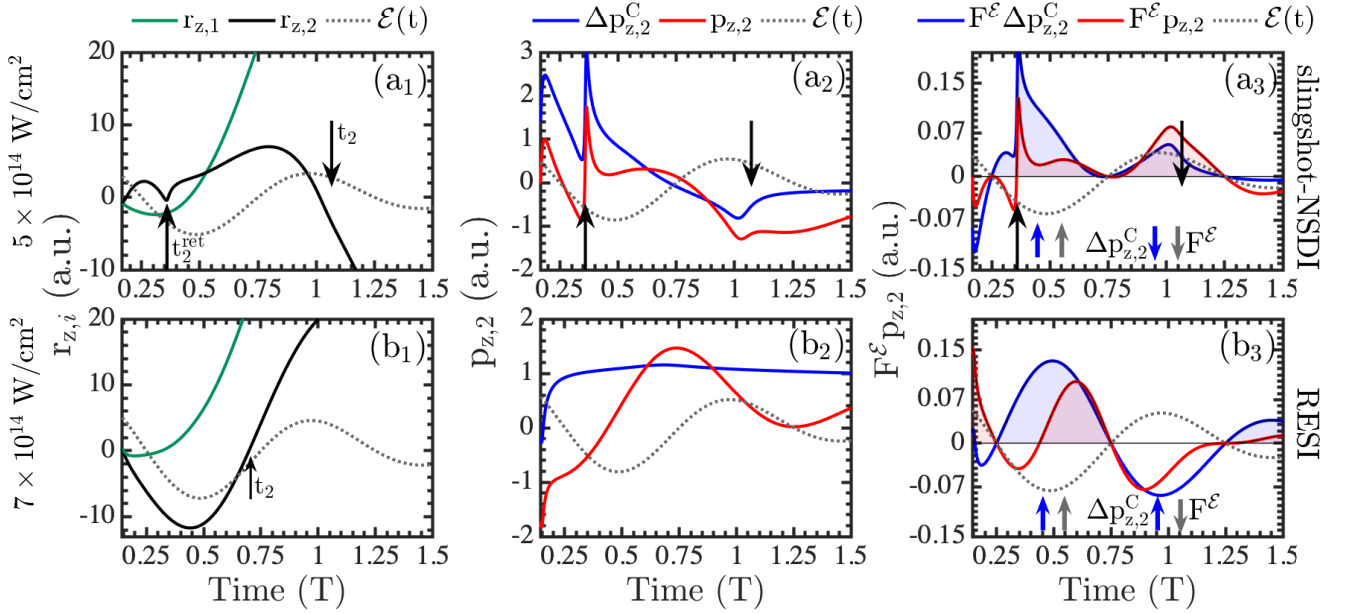}
\centering
\caption{Slingshot-NSDI at  5$\times$10$^{14}$ W/cm$^2$ (a) and RESI at  7$\times$10$^{14}$ W/cm$^2$ (b) for He driven by a near-single-cycle pulse at 400 nm. Plotted as a function of time are (1) $\mathrm{r_{z,1}}$ and  $\mathrm{r_{z,2}}$  
  (2)  $p_{z,2}$ and $\Delta p_{z,2}^{C}$ and  (3)   $F^\mathcal{E} p_{z,2}$ and $F^\mathcal{E }\Delta p_{z,2}^{C}$.  CS is enclosed by an up and down black arrow, which   represent $p_{z,2}$ being along the +$\hat{z}$-axis and -$\hat{z}$-axis, respectively, at the start and end of CS. The beginning of the time axis is $t_{rec}$. }
\label{cartoon}
\end{figure*}

  Next, we describe slingshot-NSDI  using a  representative delayed pathway event  for $\phi=0^{\circ}$ that encapsulates the main features of the new mechanism.
Below, the description of slingshot-NSDI and RESI applies to all CEPs where the re-colliding electron tunnel-ionizes from a minimum of the laser field.  A similar description applies to  CEPs where  tunnel-ionization occurs from a maximum of the  field.
  The difference is that some of  the quantities plotted  in \fig{cartoon} would be reflected with respect to the time axis. For simplicity, in \fig{cartoon}, we focus on the times after re-collision. Electron 1 (green line) ionizes  in the time interval  $\mathrm{[0,0.5]}$T with a positive momentum, mostly large as for all delayed events.

In slingshot-NSDI, electron 2 (black line) initially moves away from the nucleus along the +$\hat{z}$-axis. However, it soon returns and undergoes a close encounter with the nucleus at time t$_{2}^{\mathrm{ret}}$, see  r$_{z,2}$ as a function of time in \fig{cartoon}(a1), where r$_{z,2}$ is the z-component of the position vector ${\bf{r}}_{2}$. Indeed, r$_{z,2}$ has a very small   negative value  at time t$_{2}^{\mathrm{ret}}$. This return  of electron 2 to the nucleus  leads to a slingshot motion in the time interval t$_{2}^{\mathrm{ret}}$ to t$_{2}$, denoted by  black arrows 
in    \fig{cartoon}(a1-a3). During this  motion  electron 2 remains close to the nucleus  due to the comparable magnitude of the attractive Coulomb force and the force of the laser field. 
  Moreover, in Coulomb slingshot (CS), the momentum  of electron 2 changes significantly. It has  a large positive value at t$_{2}^{\mathrm{ret}}$, which has the same sign as the final momentum of electron 1,  to a large negative value at the time electron 2 ionizes, t$_{2}$, see red line  in \fig{cartoon}(a2). Hence, CS results in  electron 2 escaping opposite to  electron 1 along the laser field.    The effect of the nucleus has previously been addressed in the context of strongly-driven clusters \cite{Rost2008}.

For most slingshot-NSDI events electron 2 ionizes around the second extremum of the  field in the time interval [0.75, 1.25]T. To explain why this is the case, we employ   the  energy of electron 2.  Shortly after re-collision, at time  t$_{init}$=t$_{rec}$+t$_{diff}$, the repulsive force between the two electrons is significantly smaller than during re-collision. Hence, after this time, the energy of electron 2 
changes due to the work done  mainly by the   field as follows:
\begin{equation}
H(t)=\frac{p_{2}(t_{init})^{2}}{2}  -\frac{Z}{r_{2}(t_{init})}+\int_{t_{init}}^{t}F^\mathcal{E}p_{z,2}dt', 
 \label{eq:3}
  \end{equation}
where  F$^\mathcal{E}(t)=-\mathcal{E}(t)$ is the force from the laser field and  $\mathrm{F^\mathcal{E}p_{z,2}}$ is the rate of change of the energy of  electron 2.
 During CS, the close encounter of electron 2 with the nucleus at t$_{2}^{\mathrm{ret}}$   takes place past a zero of the laser  field. At this  time  both the momentum of electron 2   and the force from the laser field point along the +$\hat{z}$-axis. Roughly half a laser cycle later, in the time interval [0.75, 1.25]T, the slingshot motion is concluded with both the momentum of electron 2 and the force of the laser field pointing along the -$\hat{z}$-axis.  Thus, during  CS,  $\mathrm{F^\mathcal{E}p_{z,2}}$  is mostly positive in the first half cycle  [0.25, 0.75]T and the second one  [0.75, 1.25]T after re-collision, see red-shaded area  in \fig{cartoon}(a3).  This is the reason  electron 2 gains  sufficient energy  to ionize around the second extremum of the field after re-collision.

Next, we identify the main reason that the rate of change of the energy of electron 2 is  positive during CS. We express 
the total momentum of electron 2 as the sum of the momentum  changes  due to the interaction  with the nucleus and electron 1, $\Delta p_{z,2}^{C}$,  and with  the laser field,  $\Delta p_{z,2}^{\mathcal{E}}$ as follows:
\begin{subequations}
\begin{align}
p_{z,2}(t)&= p_{z,2}(t_{0})+\Delta p_{z,2}^{C}(t_{0}\! \rightarrow \! t) +\Delta p_{z,2}^{\mathcal{E}}(t_{0} \! \rightarrow \! t) \\
\Delta p_{z,2}^{C}(t) &=  \int_{t_{0}}^{t} \left(\frac{-Z \  r_{z,2}}{|r_2|^{3}} + \frac{r_{z,2}-r_{z,1}}{|{\bf{r}}_2 - {\bf{r}}_1|^{3}}\right)dt'\\
\Delta p_{z,2}^{\mathcal{E}}( t) &= \mathcal{A}(t) - \mathcal{A}(t_{0}). 
\label{eq:2}
\end{align}
\end{subequations}
 For delayed events,  the  repulsive force between the two electrons is roughly zero shortly after re-collision, thus,  contributing only  a  constant term  to $\Delta p_{z,2}^{C}$.
We plot the momentum change due to the nucleus as well as the total momentum of electron 2   in \fig{cartoon}(a2).
It is clear that, during CS,  the sharp change of the total momentum of electron 2      is mainly due  to the term $\Delta p_{z,2}^{C}$. Hence, the attractive force of the nucleus causes the  main change in the total momentum of electron 2. 
Combining Eqs (2) and (3), we also plot in \fig{cartoon}(a3) the contribution of the nucleus, $F^\mathcal{E }\Delta p_{z,2}^{C}$ (blue line), to the rate of change of the energy of electron 2, $\mathrm{F^\mathcal{E}p_{z,2}}$ (red line).  We find that, during CS,  the contribution  of the  nucleus   (blue-shaded area)  causes this rate to be mostly positive (red-shaded area). 
The contribution of the field,  $F^\mathcal{E}\Delta p_{z,2}^{\mathcal{E}}$ (not shown), is also mostly positive during the last part of CS.

We find that the dominant mechanism of the delayed pathway  is  slingshot-NSDI at 5$\times$10$^{14}$ W/cm$^2$ and RESI  at  7$\times$10$^{14}$ W/cm$^2$. We illustrate RESI in \fig{cartoon}{b}. In contrast to  slingshot-NSDI, in RESI the momentum change of electron 2 due to the nucleus is almost a constant after re-collision, see blue line in \fig{cartoon}(b2). Moreover, the strength of the laser field $F^\mathcal{E }$ is larger at the higher intensity. Thus, in the first half cycle,   the rate of change of the energy of electron 2, $\mathrm{F^\mathcal{E}p_{z,2}}$ in \fig{cartoon}(b3),  has larger positive values at 7$\times$10$^{14}$ W/cm$^2$ rather than at  5$\times$10$^{14}$ W/cm$^2$.  Hence, electron 2 ionizes mostly around the first extremum in RESI at 7$\times$10$^{14}$ W/cm$^2$.  The two mechanisms are compared in detail  in supplemental material \cite{sup_mat}.

 \begin{figure}[ht]
\includegraphics[scale=0.45]{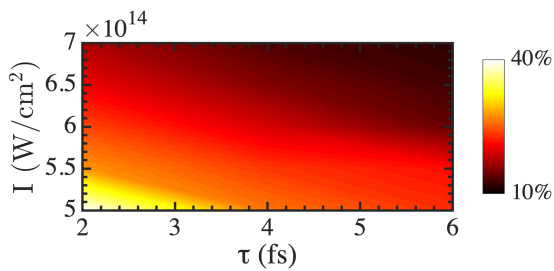}
\centering
\caption{Contribution of slingshot-NSDI to the delayed pathway as a function of laser intensity in steps of  0.5$\times$10$^{14}$ W/cm$^2$ and of pulse duration in steps of 2 fs. }
\label{fig:3}
\end{figure}

Slingshot-NSDI for He driven by a 2 fs  laser pulse at 5$\times$10$^{14}$ W/cm$^2$  and 400 nm accounts for roughly 40\% of  the delayed pathway. 
In \fig{fig:3} we demonstrate that slingshot-NSDI is a general mechanism that significantly contributes  to the delayed pathway of NSDI   for a  wide range of laser   intensities and  pulse durations for He at 400 nm. First we keep the pulse duration  constant 
and increase the intensity from 5$\times$10$^{14}$ to 7$\times$10$^{14}$ W/cm$^2$. Besides the delayed pathway, another important  pathway  is the double delayed  one at 5$\times$10$^{14}$ 
W/cm$^2$ and the direct one at    7$\times$10$^{14}$ W/cm$^2$. The contribution of slingshot-NSDI to the delayed pathway is significant for all these pulse parameters, however, it decreases with increasing intensity, see \fig{fig:3}. This   is consistent  with the increasing contribution of  RESI to the delayed pathway.

Keeping the intensity constant, we change the pulse duration  from 2 fs to 6 fs. We find that the contribution of slingshot-NSDI to the delayed pathway decreases with increasing pulse duration, see \fig{fig:3}. For a longer pulse the force from the laser field is larger at a given extremum of the field. Hence, we conjecture that it is possible that, while CS still occurs,  CS is not as pronounced and the last electron finally ionizes at subsequent half cycles also due to the contribution of the field to the rate of change of the energy of electron 2, F$^\mathcal{E}\Delta p_{z,\mathit{l}}^{\mathcal{E}}$. Moreover, a longer pulse allows for a more complicated interaction of electron 2 with the nucleus. This can result in electron 2 escaping in the same or opposite direction from the first electron. The above arguments are consistent with our finding that  in the delayed pathway the anti-correlation pattern is more  pronounced for shorter compared to longer pulses. 

For He driven at 400 nm, two experiments are most relevant to the work presented here \cite{Taylor1, Dorner2018}. In Ref. \cite{Taylor1}, 
 excellent agreement was found between experiment and  fully ab-initio quantum mechanical calculations for electron energy distributions of doubly ionized He driven by a long duration laser pulse \cite{Taylor1}. For the latter observables, using the model described here,  we  achieved excellent agreement with  fully ab-initio quantum mechanical calculations \cite{Agapi2}; the latter were  performed by the same theoretical group as in  Ref. \cite{Taylor1}. After submission of the present work, a study of a kinematically complete experiment for He driven by a long duration laser pulse at 400 nm and intensities of 3.5-5.7$\times$10$^{14}$ W/cm$^2$ was submitted and published \cite{Dorner2018}. In this latter work, the correlated electron momenta are obtained.
As the intensity increases  from 3.5$\times$10$^{14}$ W/cm$^2$  to  5.7$\times$10$^{14}$ W/cm$^2$, a transition from anti-correlated plus correlated  to mostly correlated two-electron escape is observed. We also find such a transition taking place 
 for He driven at 400 nm, as the intensity increases from 5$\times$10$^{14}$ W/cm$^2$  to  7$\times$10$^{14}$ W/cm$^2$, for  2 fs, 4 fs and 6 fs  laser pulses, see \cite{sup_mat}.
 
 To guide  experiments, we identify favorable laser parameters for observing an anti-correlated two-electron escape mostly due to slingshot-NSDI.  In the direct pathway   the two electrons escape overwhelmingly in the same direction, while in the double delayed they escape   mostly in opposite directions along the laser field. Thus, to observe slingshot-NSDI,  the contribution to NSDI of the double delayed pathway  
has to  be small. We find that for the currently considered  low  intensities, the contribution of the double delayed pathway is smaller for the shorter duration laser pulses.  We have  also shown that,
 for low intensities,  slingshot-NSDI  contributes most to the delayed pathway for short pulse durations. Given the above, it is essential that a short pulse duration is employed   for observing  slingshot-NSDI.

In conclusion,    slingshot-NSDI is an  important mechanism of the delayed pathway  and of NSDI  for a  range of intensities and pulse durations for He driven at 400 nm.  In slingshot-NSDI, following  re-collision, the electron that ionizes last undergoes a Coulomb slingshot within roughly half a laser cycle that changes sharply the direction of the electron  momentum along the  laser field. This electron has a large momentum that points in the same direction as the force from the laser field both at the start and at the end of CS.   Hence, during CS  the laser field supplies sufficient energy to this electron leading to its ionization around the second extremum of the field. The hallmark of slingshot-NSDI is that the two electrons escape in opposite directions along the laser field. 
%We also  find  (not shown) slingshot-NSDI to be  an important mechanism in NSDI of driven He at 800 nm and at a low intensity. 
We expect slingshot-NSDI to be a significant mechanism  in NSDI of  He driven at  wavelengths other than 400 nm, and  to be present in other atoms and molecules.

\begin{acknowledgments}
A. E. acknowledges  the EPSRC grant no. N031326 and the use of the computational resources of Legion at UCL. M.F.K acknowledges support by the German Research Foundation (DFG) via the Cluster of Excellence: Munich Centre for Advanced Photonics (MAP), and by the European Union (EU) via the European Research Council (ERC) grant ATTOCO.
\end{acknowledgments}

%%%
%%%

\bibliography{Resni_2018_paper}{}
\bibliographystyle{apsrev4-1}
\end{document}